\documentclass[a4paper,12pt]{article}
\usepackage{geometry}
\usepackage[cp1251]{inputenc}
\usepackage{epsfig}
\usepackage{graphicx}

\usepackage{amssymb}

\begin{document}

Astronomy Letters, 2013, Vol. 39, No. 7, pp. 01-12 \hspace{20pt} 
Printed 11 June 2013

\vspace{-8pt}

\section*{
\begin{center}
Magnetically Active Stars in Taurus--Auriga:\\ Activity and Rotation
\end{center}
}

\begin{center}
\Large K. N. Grankin
\end{center}

\begin{center}
\it{Crimean Astrophysical Observatory, Nauchny, Crimea, 98409 Ukraine}
\end{center}

\begin{center}
konstantin.grankin@rambler.ru
\end{center}

\subsection*{\center Abstract}

\begin{quote}
A sample of 70 magnetically active stars toward the Taurus–Auriga 
star-forming region has been investigated. The positions of the sample 
stars on the Rossby diagram have been analyzed. All stars are shown to 
be in the regime of a saturated dynamo, where the X-ray luminosity 
reaches its maximum and does not depend on the Rossby number. A 
correlation has been found between the lithium line equivalent width 
and the age of a solar-mass (from 0.7 to 1.2~$M_\odot$) 
pre-main-sequence star. The older the age, the smaller the Li line 
equivalent width. Analysis of the long-term photometric variability of 
these stars has shown that the photometric activity of the youngest 
stars is appreciably higher than that of the older objects from the 
sample. This result can be an indirect confirmation of the evolution of 
the magnetic field in pre-main-sequence stars from predominantly dipole 
and axisymmetric to multipole and nonaxisymmetric.\end{quote}

\vspace{15pt}

\textbf{Key words:} {\it stars -- physical properties, rotation, activity,
pre-main-sequence stars.}

\subsection*{INTRODUCTION}
\indent

Previously (Grankin 2013b), we investigated a
sample of 74 magnetically active stars toward the
Taurus–Auriga star-forming region (SFR). The
sample contains 24 well-known young pre-main-sequence
(PMS) stars and 60 candidates for
PMS stars from Wichmann et al. (1996). All sample
objects exhibit no optical and near-infrared excesses
and, consequently, have no accretion disks. Based
on accurate data on their main physical parameters
(see Tables 3 and 4 in Grankin 2013a) and published
data on their proper motions, X-ray luminosities, and
equivalent widths of the $\rm H\alpha$ and Li lines (see Table 1
in Grankin 2013b), we refined the evolutionary
status of these objects. As a result, we identified
a group of 70 objects with ages 1–40 Myr. We
showed that 50 stars from this group belong to the
Taurus–Auriga SFR with a high probability. Other
20 objects have a controversial evolutionary status
and can belong to both the Taurus–Auriga SFR and
the Gould Belt (see Table 3 in Grankin 2013b). For 50
PMS stars with known rotation periods, we analyzed
the relationship between their rotation, mass, and
age. The rotation was shown to depend on both
mass and age of young stars. We investigated the
angular momentum evolution for the sample stars
within the first 40 Myr. An active interaction between
the sample stars and their protoplanetary disks was
shown to have occurred on a time scale from 0.7 to
10 Myr. In this paper, we discuss various magnetic activity 
parameters for the sample PMS stars and
investigate the relationship between their activity and
rotation.
 
\subsection*{ACTIVITY AND ROTATION}
\indent

All of the PMS stars from our sample exhibit an
enhanced variable X-ray emission, which is evidence
for the existence of a hot corona and, hence, magnetic
activity. The key unsolved question regarding this X-ray
emission is whether an analogy exists between the
magnetic activity of PMS stars and that of the Sun.

On the Sun and on all of the stars whose internal
structure consists of a radiative core and a
convective envelope, magnetic activity is probably
produced by the so-called $\alpha\Omega$ dynamo mechanism.
This mechanism operates in a thin shell lying at the
interface between the radiative and convective zones
and is generated by the interaction between differential
rotation and convective motion (see Schrijver
and Zwaan (2000) and references therein). This hypothesis
is confirmed by the existence of magnetic
activity, which manifests itself through cool photospheric
spots, the chromospheric emission in the
calcium H and K lines and the $\rm H\alpha$ line, and the
coronal X-ray emission. In the long run, this type of
dynamo is governed by stellar rotation. As a result,
a strong activity–rotation correlation is observed for
solar-type stars with ages \textgreater100 Myr. This correlation
was first found by Skumanich (1972) and was
subsequently confirmed by numerous studies (see,
e.g., Patten and Simon 1996; Terndrup et al. 2000;
Barnes 2001). The correlation usually manifests itself
as a linear increase in activity indicators with
increasing rotation rate accompanied by activity saturation
at a high rotation rate. However, PMS stars
are known to be fully convective and, thus, cannot
provide a basis for the existence of a solar-type
dynamo. As PMS stars evolve toward the main
sequence (MS), their rotation rate changes greatly.
This can lead to a change in the magnetic field generation mechanisms
themselves and, as a consequence,
to a change in magnetic activity properties and their
relationship to stellar rotation. Thus, investigation of
the relationship between magnetic activity and rotation
for PMS stars can provide an understanding of
the fundamental changes in the physics of stars that
occur in the range of ages 1–100 Myr. In the next
sections, we will investigate the relationship between
the rotation of PMS stars in Taurus–Auriga and
various magnetic activity indicators for these stars.

The nature of the relationship between magnetic
activity and rotation can be complex, because it depends
on the stellar age, mass, internal structure,
and, possibly, interaction with the disks at early evolutionary
stages. To separate the various processes,
we will begin by analyzing the equivalent width of the
$\rm H\alpha$ line (EW($\rm H\alpha)$) as a function of the temperature
or spectral type. It should be noted that most of
the objects from our sample were identified in X-ray
surveys and the sample may be biased toward more
active objects. Therefore, when discussing activity,
we prefer to use criteria based on the upper limit of
activity in our sample and not on the lower limit,
because the latter can be shifted greatly.

\subsection*{CHROMOSPHERIC ACTIVITY}
\indent

The $\rm H\alpha$ line is commonly used as an indicator
of chromospheric activity resulting from photoionization
and collisions in a hot chromosphere. In Fig. 1,
EW($\rm H\alpha$) is plotted against the effective temperature
represented by the spectral type in our case. The
black and white symbols denote the objects with a
reliable and unreliable evolutionary status, respectively
(for details, see Grankin 2013b). The objects
classified as weak-lined T Tauri stars (WTTS) with
ages \textless10 Myr and as post T Tauri stars (PTTS)
with ages \textgreater10 Myr are marked by the circles and
squares, respectively. It can be seen from the figure
that EW($\rm H\alpha)$ is an obvious function of the spectral
type. Whereas the $\rm H\alpha$ line in G-type stars is in
absorption (EW($\rm H\alpha) \sim 2$\AA), $\rm H\alpha$ in K1–K4 stars
exhibits a gradual transition to an emission state.
For stars later than K5, $\rm H\alpha$ is always in emission.
The scatter of EW($\rm H\alpha)$ increases and EW($\rm H\alpha)$ for
M-type stars lie within the range from 0 to $-7$~\AA.
The dramatic change in EW($\rm H\alpha)$ with spectral type
reflects not only the change in chromospheric activity
but also the additional effects related to a reduction
in the continuum level with decreasing stellar luminosity
and to a change in photospheric absorption
in $\rm H\alpha$, which is zero for M dwarfs and increases
toward earlier spectral types. The combined effects of
a reduction in the continuum level and photospheric
absorption in $\rm H\alpha$ were estimated from spectroscopic
observations of standard nonactive stars by Scholz
et al. (2007). The linear approximation of EW($\rm H\alpha)$
as a function of the spectral type is indicated in Fig. 1
by the dash–dotted line. This line coincides with the
dependence of EW($\rm H\alpha)$ on spectral type for stars in
the Hyades and for field stars of various spectral types
(for details, see Scholz et al. 2007). Thus, this line is
an estimate for the purely photospheric contribution
to EW($\rm H\alpha)$.

\begin{figure}[ht]
\epsfxsize=12cm
\vspace{0.6cm}
\hspace{2cm}\epsffile{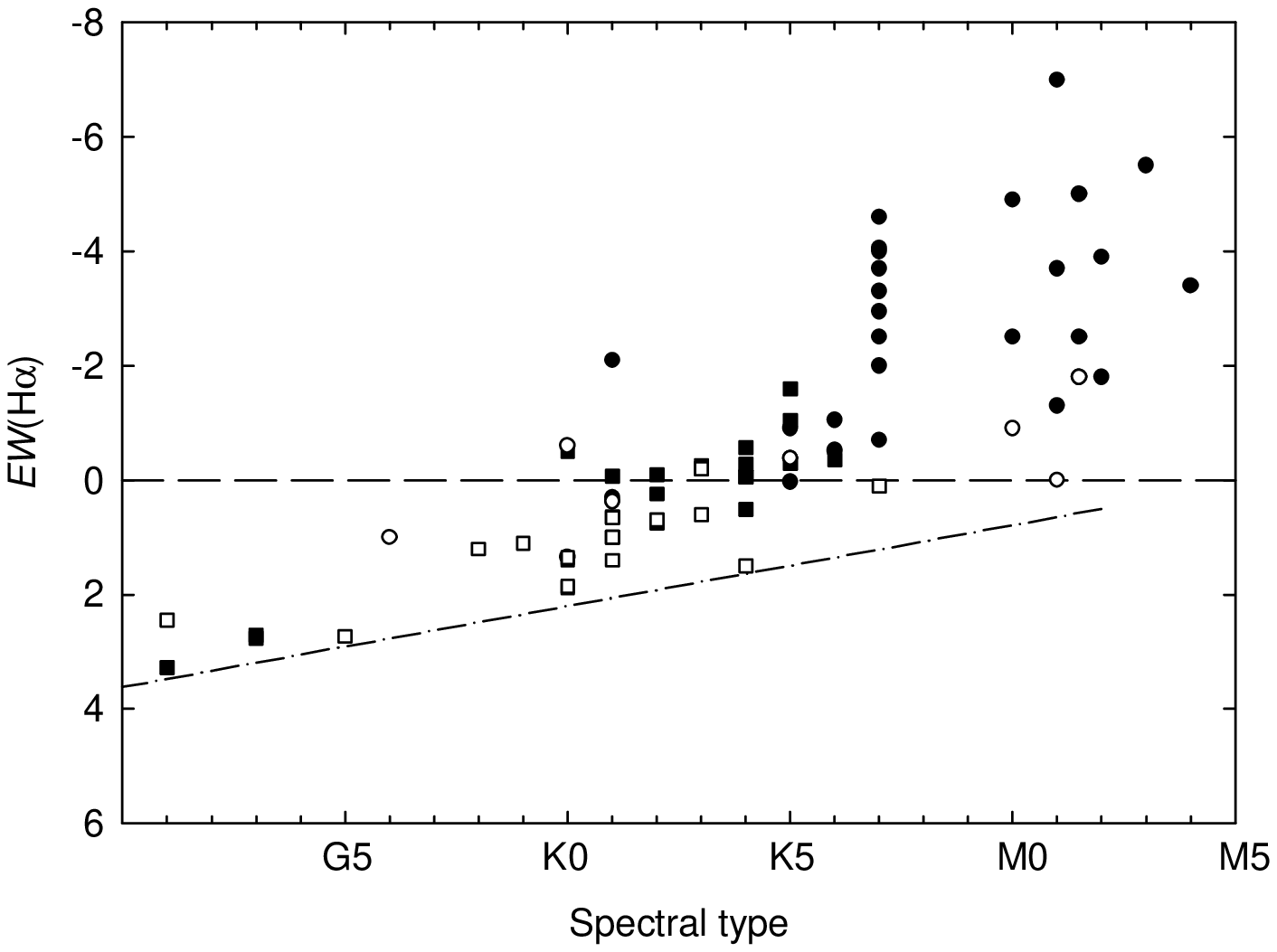}
\caption{\rm \footnotesize {EW($\rm H\alpha)$ versus spectral type. The 
dash–dotted line indicates the upper limit of EW($\rm H\alpha)$ for 
nonactive field stars. The dashed line corresponds to zero for 
EW($\rm H\alpha)$. The black and white symbols denote the objects with 
a reliable and unreliable evolutionary status, respectively. The 
objects classified as WTTS (with ages \textless10 Myr) and PTTS (with 
ages \textgreater10 Myr) are marked by the circles and squares, respectively.}}
\end{figure}

It can be seen from the figure that the dash–dotted
line is the lower envelope for the stars of our sample:
almost all of the PMS stars in Taurus–Auriga lie
above this line, except for seven objects that exhibit
no measurable chromospheric activity. Thus, most
of the stars from our sample are chromospherically
active objects. The maximum activity level increases
rapidly for K7–M4 stars.

\subsection*{X-RAY ACTIVITY}
\indent

First of all, we investigated the possible relationship
between the rotation period ($P_{\rm rot}$) and various
X-ray activity parameters of PMS stars: the X-ray
luminosity ($L_{\rm X}$), the X-ray surface flux ($F_{\rm X}$), and
the X-ray luminosity excess defined as the X-ray to
bolometric luminosity ratio ($L_{\rm X}/L_{\rm bol}$). To calculate
these X-ray activity parameters, we used data from
our two previous papers (Grankin 2013a, 2013b).
Figure 2 shows the corresponding graphs on a logarithmic
scale. It can be seen from the figure that a
very weak correlation is traceable between $P_{\rm rot}$ and
$L_{\rm X}$ (see Fig. 2a). We failed to find any correlation
between the rotation period and $L_{\rm X}/L_{\rm bol}$ (see
Fig. 2b). The most significant correlation is observed
between $P_{\rm rot}$ and $F_{\rm X}$ (see Fig. 2c). Nevertheless,
it would be unreasonable to assert that there exists
an unambiguous correlation between these activity
parameters of PMS stars and their rotation period
because of the small correlation coefficients.

\begin{figure}[ht]
\epsfxsize=8.5cm
\vspace{0.6cm}
\hspace{4cm}\epsffile{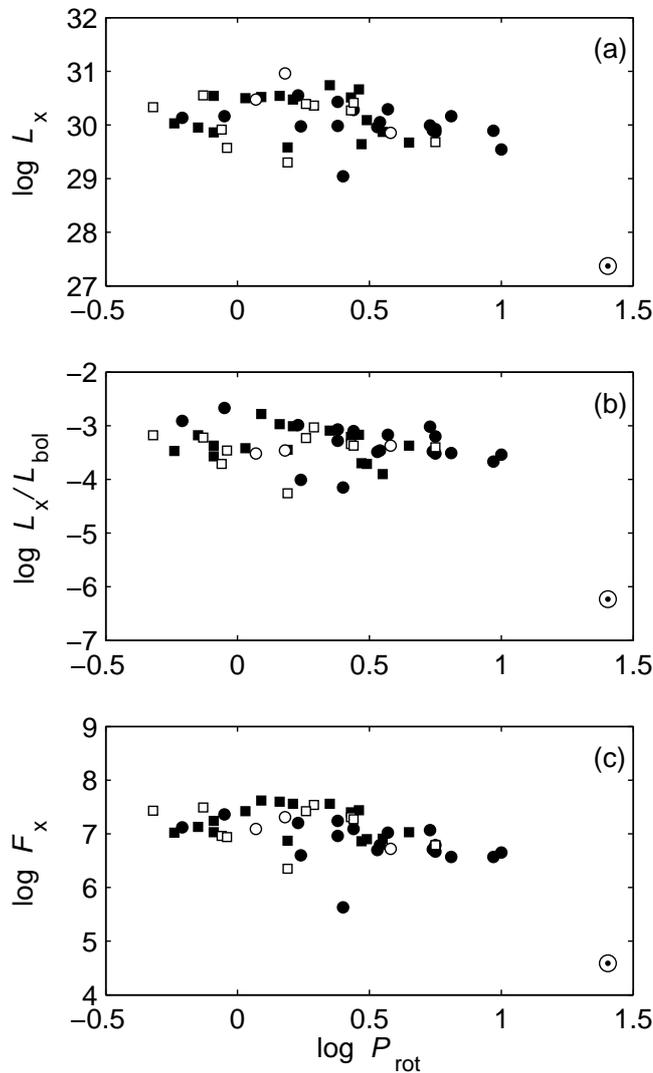}
\caption{\rm \footnotesize {Relationship between rotation period and 
X-ray activity parameters: the X-ray luminosity $L_{\rm X}$ (a), the
ratio $L_{\rm X}/L_{\rm bol}$ (b), and the X-ray surface flux 
$F_{\rm X}$ (c). The position of the Sun is denoted by the corresponding
symbol. The designations of the objects are the same as those in Fig. 1.}}
\end{figure}

We attempted to investigate the possible correlation
between the X-ray activity of PMS stars and
their angular momentum, which is a no less informative
characteristic of stellar rotation. The angular
momentum is $J=I\omega$, where $I$ is the moment of
inertia and $\omega$ is the angular velocity of the star. The
angular velocity of a star is easy to calculate if its
rotation period is known: $\omega = 2\pi/P$. The moment
of inertia of a star can be determined from the formula
$I = M(kR)^2$, where $M$ is the stellar mass, $R$ is the
stellar radius, and $k$ is the radius of inertia dependent
on the rotation period and shape of the star. Thus,
the angular momentum of a star depends on three
parameters: its mass, radius, and rotation period. To
reduce the number of unknown parameters, we can
introduce some normalized angular momentum:

$$
j=\frac{J}{M}=\frac{I\omega}{M}= \frac{2\pi M(kR)^2}{MP}=
2\pi k^{2}\frac{R^2}{P}. 
$$

For a spherically symmetric star, $k=\sqrt{2/3}$ and
the normalized angular momentum can be calculated
from the formula $j=\frac{4\pi}{3}\frac{R^2}{P}$. For the Sun,
$P$ = 25 days, $R = 6.96\times10^{10}$ cm, and the angular
momentum $j_\odot = 9.39\times10^{15} \rm cm^{2} s^{-1}$ .

\begin{figure}[ht]
\epsfxsize=8.5cm
\vspace{0.6cm}
\hspace{4cm}\epsffile{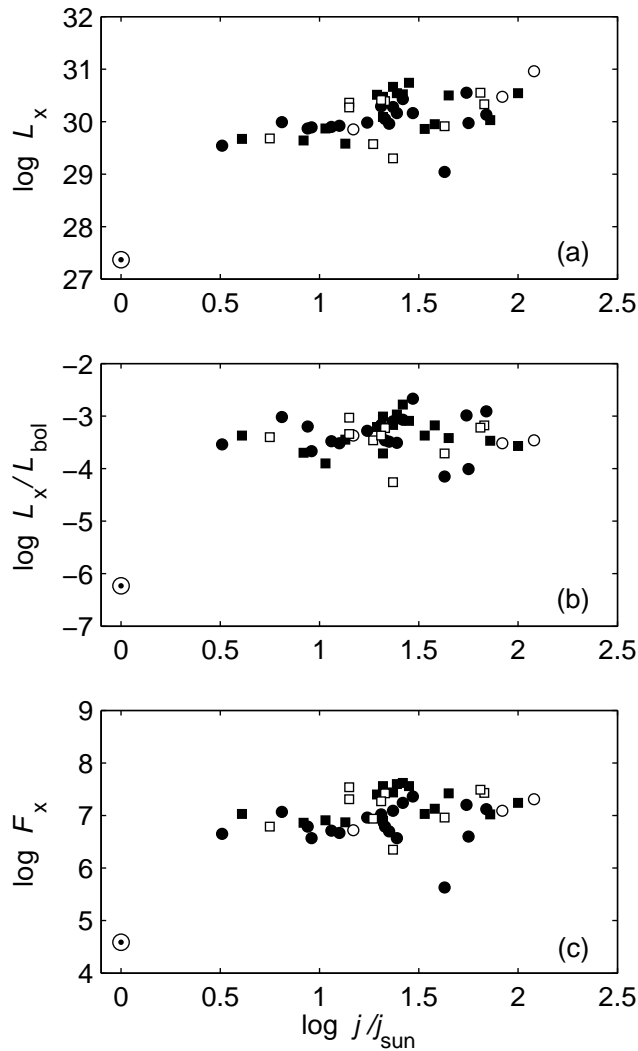}
\caption{\rm \footnotesize {Relationship between angular momentum and
various X-ray activity parameters. The sign of the Sun
in the lower left corners marks its position. The designations
are the same as those in Fig. 1.}}
\end{figure}

We calculated the ratio $j/j_\odot$ for all of the stars
from our sample with known rotation periods. Based
on these data, we plotted the normalized angular momentum(
$j/j_\odot$) against various X-ray activity parameters
of PMS stars: $L_{\rm X}$  (Fig. 3a), $L_{\rm X}/L_{\rm bol}$  
(Fig. 3b), and $F_{\rm X}$ (Fig.~3c). It can be seen from Fig. 3 
that there is no significant correlation between the various X-ray
activity parameters and angular momentum, as in
the case with the rotation period.

\subsection*{ROSSBY DIAGRAM}
\indent

The chromospheric and coronal activity of stars is
known to be related to their rotation and to the depth
of the convective zone or the convective turnover time
($\rm \tau_c$). The existence of such a relationship is in good
agreement with qualitative predictions of the $\alpha\Omega$ dynamo
theory explaining the generation of a magnetic
field. The Rossby diagram is one of the best tools for
demonstrating the existence of a relationship between
magnetic field generation and stellar activity. As a
rule, the Rossby diagram displays the relationship
between some stellar activity indicator and the Rossby
number ($R_0$) calculated from the formula 
$R_{0} = P_{\rm rot}/\tau_c$. In particular, previous studies have shown
that the slowly rotating stars in the Hyades cluster
and most of the dwarf field stars exhibit a decrease
in log$(L_{\rm X}/L_{\rm bol})$ with increasing Rossby number. In
contrast, the rapidly rotating field stars and G and K
dwarfs in the Pleiades and $\alpha$~Persei clusters exhibit
no obvious relationship between log$(L_{\rm X}/L_{\rm bol})$ and the
Rossby number (Hempelmann et al. 1995; Patten and
Simon 1996; Randich et al. 1996; Queloz et al. 1998).
Subsequently, Pizzolato et al. (2003) showed that
the relationship between the X-ray luminosity and
rotation period of a star could be roughly described by
a power law irrespective of its mass and spectral type.
Thus, stellar rotation dominates over convection for
slowly rotating solar-type stars. At the same time,
the X-ray luminosity of rapidly rotating stars depends
only on $L_{\rm bol}$ and, consequently, depends on stellar-structure
characteristics.

Recent studies have shown that for slowly rotating
stars there is a tendency for $L_{\rm X}/L_{\rm bol}$ to grow with
increasing rotation velocity up to $\backsim$15 km $\rm s^{-1}$, while
stars with higher velocities have approximately the same 
$L_{\rm X}/L_{\rm bol}$ near the saturation level 
($L_{\rm X}/L_{\rm bol}\backsim 10^{-3}$), with this saturation limit 
being observed for stars in a wide range of spectral types, from G to M.
Thus, the most active stars exhibit a maximum X-ray
luminosity at a level of $L_{\rm X}/L_{\rm bol}\backsim 10^{-3}$, which does
not depend on the rotation velocity. This phenomenon
was called the saturated dynamo effect.

In Fig. 4, log$(L_{\rm X}/L_{\rm bol})$ is plotted against the
Rossby number for all of the stars from our sample
with known rotation periods. We estimated the
convective turnover time $\rm \tau_c$ from an empirical relation
given in Noyes et al. (1984):

$$
\log \tau_c =
\left\{
\begin{array}{lcl}
1.362 - 0.166x+0.025x^2-5.323x^3,~x>0,\\
1.362-0.14x,~x<0,\\
\end{array}
\right.
$$\\
where $x=1-(B-V)$. We used the extinction-corrected
color index as $(B-V)$. To compare the activity of PMS 
stars with the activity of other solar-type
stars, we showed the positions of the MS stars
and the stars from the Hyades, Pleiades, IC 2391,
and IC 2602 open clusters investigated by Pizzolato
et al. (2003) in Fig. 4.

It can be seen from Fig.~4 that the X-ray luminosity
excess ($L_{\rm X}/L_{\rm bol}$) for the sample of active stars
from Pizzolato et al. (2003) increases with decreasing
Rossby number ($R_0$). However, the increase in
$L_{\rm X}/L_{\rm bol}$ ceases at a level of about 
log($L_{\rm X}/L_{\rm bol})=-3$, when $R_0$ 
reaches $ \backsim 0.28-0.1$ (log$R_0 = -0.56--0.98$). 
From this time on, the so-called saturation
regime is observed, where the X-ray luminosity
excess reaches its maximum values and ceases to
depend on $R_0$.

All PMS stars from our sample exhibit the same
$L_{\rm X}/L_{\rm bol}$ and $R_0$ as the stars from the IC~2602 and
Pleiades open clusters with ages within the range
30–100 Myr. In other words, the X-ray activity of
PMS stars in the Taurus–Auriga SFR closely coincides
with that of the cluster stars in the regime
of saturated activity. It should be noted that there
are slightly more stars with small log$R_0$ in the range
from --1.7 to --2.1 in the Pleiades. Since the stars
of our sample are located in the zone of a saturated
dynamo, the fact that we failed to find an unequivocal
correlation between the rotation period and various
X-ray activity parameters (see the previous section)
becomes explainable.

\begin{figure}[ht]
\epsfxsize=13cm
\vspace{0.6cm}
\hspace{2.5cm}\epsffile{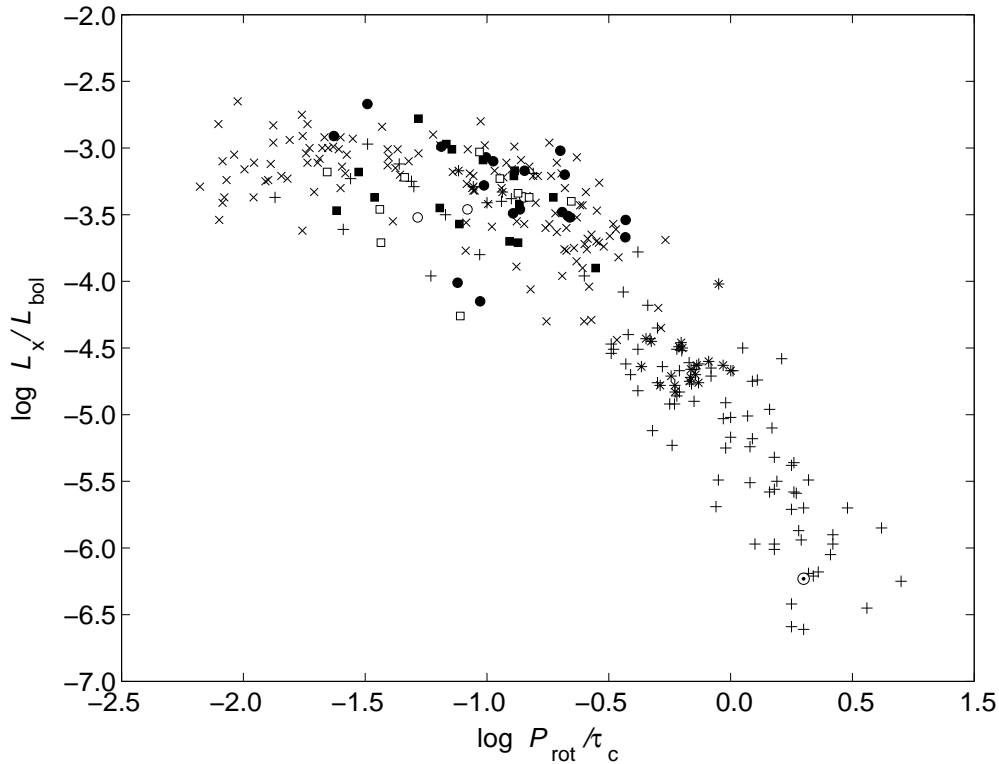}
\caption{\rm \footnotesize {$L_{\rm X}/L_{\rm bol}$  versus Rossby 
number. The crosses are stars from the Pleiades, IC 2391, and IC 2602 
open clusters; the asterisks are Hyades stars; the pluses are MS dwarfs 
from Pizzolato et al. (2003). The position of the Sun at the maximum of
the activity cycle is also marked. The designations of PMS objects are 
the same as those in Fig. 1.}}
\end{figure}

\subsection*{PHOTOSPHERIC ACTIVITY}
\indent

In previous sections, we have pointed out that the
magnetic activity of young solar-type stars manifests
itself through the chromospheric emission in the calcium
H and K lines and the $\rm H\alpha$ line or through the
coronal X-ray emission. In addition, the magnetic
activity can also manifest itself through the maximum
photometric variability amplitude in the optical spectral
range ($\Delta V_{\rm max}$). Indeed, the maximum photometric
variability amplitude depends primarily on the
degree of nonuniformity in the distribution of spotted
regions over the stellar surface and, consequently,
on the total surface magnetic flux. Before investigating
the possible relationship between $\Delta V_{\rm max}$ and
various rotation parameters, we analyzed the possible
correlations of $\Delta V_{\rm max}$ with such parameters as the
spectral type of a star and its age. In Fig. 5a, the
maximum variability amplitude is plotted against the
spectral type. It can be seen from the figure that the
maximum amplitude gradually increases from earlier
spectral types to later ones and reaches its maximum 
near a spectral type K7–M1. This effect can be due
to a change in the contrast of dark spots against the
background of the photosphere. The results of modeling
this effect are represented by the solid line. It can
be clearly seen that the amplitude of the periodicity
increases from relatively early spectral types to later
ones. It should be noted that our sample contains
five most active WTTS whose maximum amplitudes
are considerably larger than those of the remaining
PMS stars: LkCa~4, LkCa~7, V827~Tau, V830~Tau,
and TAP~41. These stars are separated from the
main group by the horizontal dashed line at a level of
$0.^m35$. We excluded these objects from the subsequent
statistical analysis and will discuss their properties
separately.

\begin{figure}[h]
\epsfxsize=13.0cm
\vspace{0.6cm}
\hspace{2.0cm}\epsffile{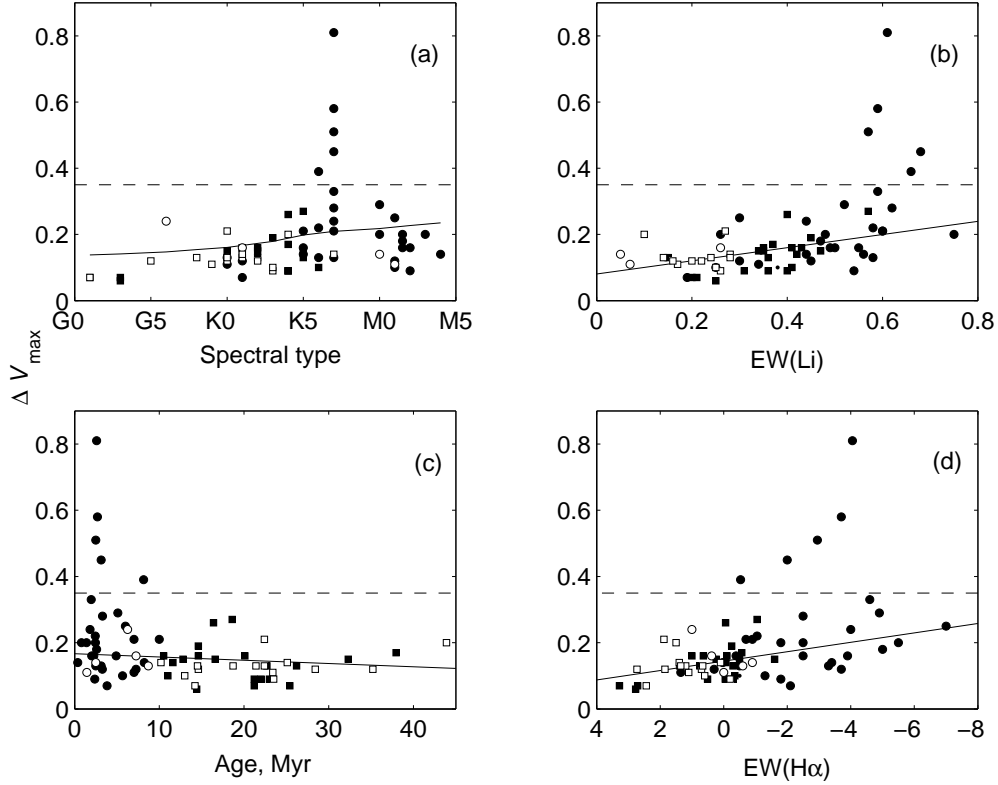}
\caption{\rm \footnotesize {Maximum photometric variability amplitude 
versus spectral type (a), Li line equivalent width (b), age (c), and 
$\rm H\alpha$ equivalent width (d). The designations of PMS objects are 
the same as those in Fig. 1.}}
\end{figure}

In Fig. 5b, $\Delta V_{\rm max}$ is plotted against EW(Li). We
found a weak correlation between $\Delta V_{\rm max}$  and EW(Li)
with a correlation coefficient k = 0.39. The maximum
photometric variability amplitude increases with increasing
lithium line equivalent width. The existence
of such a positive correlation between $\Delta V_{\rm max}$  and
EW(Li) may reflect the fact that younger PMS stars
are simultaneously also more active objects. For
example, it is easy to notice that the five most active
WTTS lying above the dashed line have values of
EW(Li) that are among the largest ones. This result
is quite intriguing, because the presence of a lithium
absorption line is considered primarily as a signature
of youth and not as a signature of stellar activity.

In Fig. 5c, the maximum photometric variability
amplitude is plotted against the age of PMS stars.
The solid line is a linear approximation for all of the
stars lying below the dashed line ($\Delta V_{\rm max}<0.^m35$). A
weak correlation between the maximum photometric
variability amplitude and age with a correlation
coefficient k = 0.45 is noticeable. The maximum
amplitude decreases with increasing stellar age. The
four stars exhibiting the largest variability amplitudes
and lying above the horizontal dashed line have ages
$2.5-3.1$ Myr.

In Fig. 5d, $\Delta V_{\rm max}$ is plotted against $\rm EW(H\alpha)$.
The solid line is a linear approximation for all of the stars lying 
below the horizontal dashed line ($\Delta V_{\rm max}<0.^m35$). It can 
be seen from the figure that there is a clear correlation between 
$\Delta V_{\rm max}$ and $\rm EW(H\alpha)$ with a correlation 
coefficient k = 0.46. As the $\rm H\alpha$ line passes from an 
absorption state to an emission one, the maximum photometric 
variability amplitude increases monotonically. This result confirms our 
assumption that the maximum photometric variability
amplitude can be used as an indicator of photospheric
activity, while the $\rm H\alpha$ line is an indicator of chromospheric
activity for PMS stars.

In the previous section, we showed that the
PMS stars in Taurus–Auriga are in the regime of
a saturated dynamo, where the X-ray flux reaches
saturation and ceases to depend on the rotation rate.
Since the X-ray flux is related to the number of active
regions on the stellar surface, we can assume that
the active regions should cover almost the entire
stellar surface in the regime of a saturated dynamo.
In that case, we may expect maximum photometric
variability amplitudes for PMS stars, of course, only
when the active regions are distributed over the
stellar surface highly nonuniformly. Therefore, it is
interesting to investigate the relationship between
$\Delta V_{\rm max}$ and such stellar rotation parameters as the
rotation period, the Rossby number, the equatorial
rotation velocity, and the angular momentum. For
this purpose, we constructed the corresponding
dependences but failed to find a clear correlation
between $\Delta V_{\rm max}$ and the rotation parameters listed
above. This result confirms our conclusion that
the PMS stars are in a state of saturated dynamo.
Regarding the five most active stars that exhibit
the largest variability amplitudes ($\Delta V_{\rm max}> 0.^m35$),
it should be noted that they have moderate rotation
velocities ($v_{\rm eq} = 10-30$ km $\rm s^{-1}$) and are in the region
of the transition into the zone of a saturated dynamo
that corresponds to Rossby numbers in the range
log$(P_{\rm rot}/\tau_c) = -0.56 - -0.98$.

\vspace{-2pt}
\subsection*{Li EVOLUTION DURING THE PMS STAGE}
\indent

Lithium, just as other light elements, such as
beryllium and boron, is burnt out in thermonuclear
reactions at relatively low temperatures in the stellar
interior ($(2.5-3.0)\times 10^6$ K). In the case of initial
evolution of low-mass ($M < 1.2~M_\odot$) stars, efficient
mixing can rapidly transport a lithium-depleted material
from the central regions of a PMS star to its
photosphere. For this reason, measurements of the
photospheric Li abundance provide one of the few
means for probing the stellar interior and are a sensitive
test of evolution models for PMS stars. Understanding
the Li depletion mechanisms at the stage of
PMS stars also makes it possible to estimate the ages
of young stars.

A large number of observational and theoretical
works were devoted to understanding the initial
abundance of Li and its subsequent PMS evolution
(see the recent review by Jeffries 2006). According
to classical models, the photospheric depletion of
Li begins near 2 Myr for a star with a mass of
1 $M_\odot$ and should end in an age of about 15 Myr.
This window moves toward older ages for stars with
lower masses. However, the degree of Li depletion
depends very strongly on mass, convection efficiency,
opacity, metallicity, and other model parameters.
Thus, the amount of photospheric Li can serve as a
characteristic of the youth of a PMS star. Nevertheless,
numerous observations of the photospheric
lithium abundance in hundreds of young stars in
open clusters suggest that the degree of Li depletion
in these stars is considerably smaller than what is
predicted by standard models. In addition, the K-type
stars in clusters with ages of less than 250 Myr are
characterized by a significant dispersion in Li abundance
whose cause is not yet completely understood
(Jeffries 2006).

This and other puzzling peculiarities that are unexpected
within the framework of standard models
suggest that Li depletion is governed not only by
convection and that there exist other, unknown processes
that have not been included in the classical
theory. In recent years, several nonstandard models
explaining the physics of the possible processes leading
to Li depletion have been proposed. However,
the mechanisms governing Li depletion still remain
poorly studied. Additional observational constraints
for present-day models are badly needed for further
progress in understanding the evolution of Li during
the PMS stage.

That is why we attempted to reveal any correlations
or relationships between the Li equivalent width
and other physical parameters of the PMS stars from
our sample. Finding such correlations can shed light
on the problem of Li depletion at the PMS evolutionary
phase of young stars. In the previous section,
we pointed out that there exists a weak positive correlation
between EW(Li) and $\Delta V_{\rm max}$ for PMS stars.
At the same time, we failed to find any correlation
between EW(Li) and the X-ray luminosity.

Below, we investigate the possible correlation between
EW(Li) and such parameters of PMS stars as
the theoretical age ($t$) and the rotation period ($P_{\rm rot}$).

If the entire sample of stars is considered as a single
group, then no correlation is observed between
EW(Li) and $t$. It should be noted that our sample
includes PMS stars with quite different physical parameters.
For example, the masses of the PMS stars
from our sample lie within the range 0.26 -- 2.2 $M_\odot$.
Since the degree of Li depletion depends very strongly
on mass, we attempted to find a possible correlation
between EW(Li), age, and rotation period for stars
with masses fairly close to the solar mass (in the range
0.7 -- 1.2\,$M_\odot$). In Fig. 6, EW(Li) is plotted against $t$
(a) and $P_{\rm rot}$ (b).

\begin{figure}[h]
\epsfxsize=14cm
\vspace{0.6cm}
\hspace{1.5cm}\epsffile{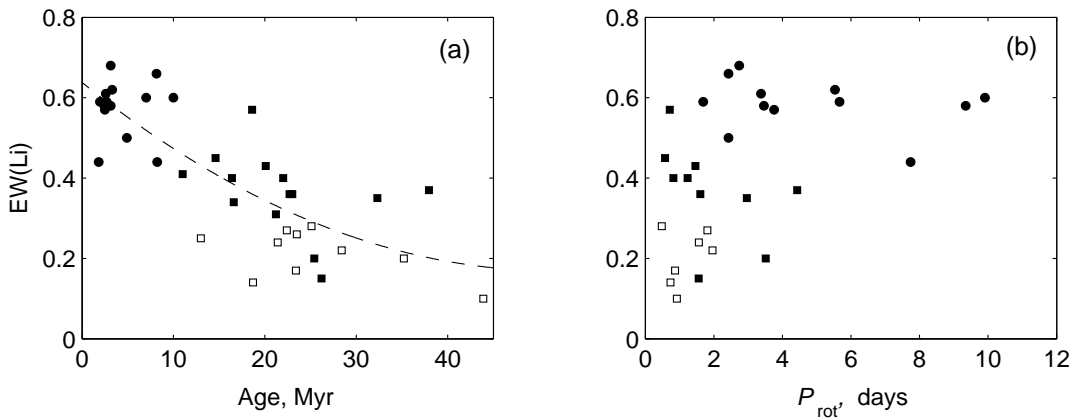}
\caption{\rm \footnotesize {EW(Li) versus age (a) and rotation period 
(b). The designations are the same as those in Fig. 1.}}
\end{figure}

For solar-mass stars, there exists a statistically
significant correlation between EW(Li) and $t$ with a
correlation coefficient of 0.68. The older the age, the
smaller the equivalent width EW(Li). The stars with
ages $\backsim2-3$ Myr have maximum values of EW(Li)
($\backsim 0.58$\,\AA). In contrast, the stars with ages older than
30 Myr exhibit minimum values of EW(Li) (about
0.20\,\AA). This result is in excellent agreement with
the predictions of the classical models explaining the
evolution of the atmospheric Li abundance during the
PMS evolutionary stage of solar-mass stars.

If the entire sample of stars is considered as a
single group, then no correlation is observed between
EW(Li) and $P_{\rm rot}$. It can also be noted that
the stars with rotation periods longer than 5 days
have relatively broad Li absorption lines 
(EW(Li)\textgreater 0.4\,\AA). In contrast, the stars with periods 
shorter than 5 days have various values of EW(Li), from 0.2 to 
0.7\,\AA. If each subgroup of stars is considered separately, then the 
following can be identified: (1) the reliable WTTS (black circles) have
EW(Li) $\backsim 0.6\,$\AA\ for the entire range of rotation
periods from 0.5 to 10 days; (2) the reliable PTTS
(black squares) with ages older than 10 Myr have
 $\rm EW(Li) \backsim 0.4\,$\AA\ for their range of rotation periods
from 0.5 to 5 days; (3) the PTTS with an unreliable
evolutionary status rotate more rapidly ($P_{\rm rot}$ in
the range from 0.5 to 2 days) and have $\rm EW(Li) \backsim 0.2\,$\AA. 
In other words, the same dependence of
the lithium line equivalent width on age, but not on
rotation period, is observed.

\subsection*{PROPERTIES OF THE MOST ACTIVE PMS STARS}
\indent

Several most active stars that exhibit the record maximum photometric 
variability amplitudes deserve particular attention: LkCa~4 ($0^m.81$), 
LkCa~7 ($0^m.58$), V827~Tau [TAP~42] ($0^m.51$), V830~Tau ($0^m.45$), 
and V1075~Tau [TAP~41] ($0^m.39$). It should be noted that there are 
two more objects that exhibit large variability amplitudes: V410~Tau 
($0^m.63$) and V836~Tau ($0^m.62$). However, we do not discuss their 
properties here, because we failed to determine reliable luminosities,
radii, masses, and ages for them. Such large amplitudes of light 
variations can be due to the existence of very large and extended 
spotted regions in the photospheres of these stars; these extended 
spotted regions must be distributed over the surface highly 
nonuniformly, otherwise the very large amplitudes of periodic light 
variations reaching $0^m.4-0^m.8$ cannot be explained.

A previous analysis of long-term photometric observations for a sample 
of well-known WTTS from the Taurus–Auriga SFR showed that some of these
objects exhibit stability of the phase of minimum light 
($\varphi_{\rm min}$) over several years of observations (see, e.g., 
Grankin et al. (2008) and references therein). Only seven stars from 
the entire sample of known WTTS show stability of $\varphi_{\rm min}$ 
in the interval from 5 to 19 years: LkCa~4, LkCa~7, V819~Tau, V827~Tau,
V830~Tau, V836~Tau, and V410~Tau. Such long-term stability of 
$\varphi_{\rm min}$ can be due to the existence of the so-called active 
longitudes at which short-lived groups of spots are located (Grankin et 
al. 1995). Similar long-lived active regions are known on the Sun and 
some of the RS CVn binary stars.

It should be noted that almost all of these stars
enter into the list of seven most active objects that
exhibit the largest photometric variability amplitudes
(see above). Thus, the stability of the phase of minimum
light for these objects is somehow related to the
existence of very large variability amplitudes. Given
the unusual photometric properties of these active
stars, we decided to discuss their main physical parameters
in more detail.

First, all of the most active stars with known parameters have very 
similar spectral types in the range K6–K7 and, hence, almost identical 
surface temperatures. Second, the rotation periods of these active 
stars lie within a fairly narrow range, from 2.4 to 5.7 days. Third, 
the radii of these stars are fairly close and lie within the range from 
1.30 to 1.75\,$R_\odot$. Fourth, analysis of their positions on the 
Hertzsprung–Russell diagram showed that their masses also have close 
values (from 0.74 to 0.92\,$M_\odot$), while their ages lie within the 
range from 2.5 to 8.2 Myr.

Apart from the similarity of the main parameters of these stars noted 
above, it should be noted that they exhibit the broadest lithium lines 
with $\rm EW(Li) > 0.57$\,\AA. This fact confirms our assumption that 
the lithium line equivalent width can be used as an activity and youth 
criterion for stars, because there exists a significant correlation 
between EW(Li), the mean photometric variability amplitude, and the age
of PMS stars (see Figs. 5b and 6a).

We limited our sample of stars in mass and considered only those 
objects whose masses were close to the solar mass (within the range 
from 0.7 to 1.2\,$M_\odot$). In this case, all five most active stars 
exhibit the strongest $\rm H\alpha$ emission lines and are among the
youngest ones, with ages from 2.5 to 3 Myr, except TAP~41
whose age is estimated to be 8 Myr.

Thus, if the subgroup of solar-mass PMS stars is considered, then it 
can be asserted that the most active and youngest stars with ages of no 
more than 8 Myr have the largest photometric variability amplitudes
(reaching $0^m.39-0^m.81$), show $\rm H\alpha$ emission in the range 
from --0.5 to --4.0\,\AA, and exhibit the most stable phase light 
curves and the strongest lithium absorption line (EW(Li)
\textgreater 0.57\,\AA).

\subsection*{EVOLUTION OF THE PHASE LIGHT CURVES}
\indent

The evolution of the phase light curves for the most active PMS stars 
was analyzed in detail by Grankin et al. (2008). Here, we will point 
out the most interesting results. Despite the stability of the phase 
light curves in the sense of stability of the phase of minimum light, 
all active stars exhibit significant changes in the amplitude and shape 
of the phase light curve from season to season. For example, the amplitude
for LkCa~7 changes from $0^m.33$ to $0^m.58$. The most symmetric 
(relative to $\varphi_{\rm min}$) phase light curve, as a rule, 
corresponds to the season with a maximum amplitude. Conversely, the 
most asymmetric phase light curves correspond to minimum amplitudes of
light variations.

Whereas many of the stars show gradual changes in the amplitude and 
shape of the light curve from season to season (for example, LkCa~4 and 
LkCa~7), there are examples of a completely different behavior. The 
amplitude of periodic light variations can change noticeably by a few 
tenths of a magnitude during one season, as in the case of TAP~41.

Although the amplitude of the light curve can change noticeably, the 
mean brightness level is essentially constant. Simple simulations 
showed that the stability of the mean brightness level from season
to season suggests that the total number of spots on the surfaces of 
active stars changes much less than their distribution over the stellar 
surface. In other words, the decrease in the amplitude of the periodicity
is attributable not to a decrease in the total area of the spots (in 
this case, the mean brightness level should increase) but to a more 
uniform distribution of the spots over the stellar surface. However, it 
should be noted that some of the stars exhibit noticeable changes in 
the mean brightness level from season to season (V819~Tau, V827~Tau, 
V836~Tau, and VY~Tau).

All the noted peculiarities of the evolution of the phase light curves 
concerned the most active and youngest stars of our sample. However, 
the overwhelming majority of sample stars show a slightly different
photometric behavior that was not discussed in previous papers. In 
particular, many of the PMS stars exhibit modest variability amplitudes 
that do not exceed $0^m.1-0^m.2$. In addition, periodic light variations
are observed much more rarely than in the most active PMS stars whose 
properties have been discussed above. The differences in photometric 
behavior between the most active PMS stars and the remaining sample 
stars are presented in Fig. 7. Figures 7a and 7b show the seasonal 
phase light curves, respectively, for V819~Tau, one of the most active 
and youngest WTTS with an age of 3.3 Myr, and for W62 (RXJ0452.5+1730), 
a reliable PTTS with an age of 22 Myr. It can be seen from the figure 
that clear periodic light variations in W62 were detected only in two 
of the six seasons: in 2001 and 2004. In contrast, periodic light 
variations in V819~Tau were observed during each observing season. The 
phase of minimum light remained stable over 6 years (from 1999 to 2004).

To quantitatively characterize the frequency of
occurrence of a periodicity, we used a simple parameter,
$f=N_{\rm p}/N_{\rm s}$, where $N_{\rm p}$ is the number of
observing seasons with periodic light variations and
$N_{\rm s}$ is the total number of observing seasons (see
Grankin 2013a). For example, the most active
PMS stars exhibit periodic light variations virtually
during each observing season, i.e., the frequency of
occurrence of a periodicity is $f = 1$.

The less active PMS stars show periodic light
variations with a mean frequency $\backsim$0.5, i.e., periodic
light variations were detected in half of the observing
seasons. Since most of the stars from our sample
are not so young and active, we investigated the
relationship between the frequency of occurrence of
a periodicity and such parameters of PMS stars as
EW(Li) (Fig. 8a) and the age (Fig. 8b). It can be
seen from Fig. 8a that the frequency of occurrence of
a periodicity increases with increasing EW(Li). The
dependence of $f$ on age (Fig. 8b) shows that the frequency
of occurrence of a periodicity is at a maximum
for the youngest stars and gradually decreases for
older objects.

\begin{figure}[h]
\epsfxsize=13cm
\vspace{0.5cm}
\hspace{1cm}
\epsffile{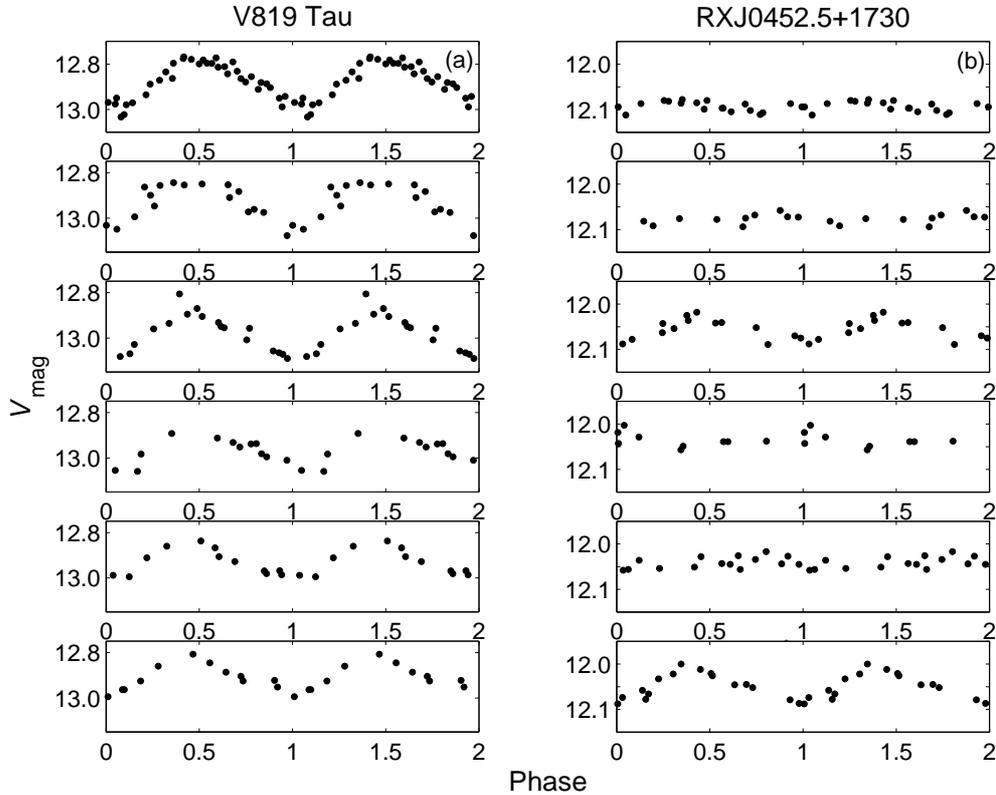}
\caption{\rm \footnotesize {Phase light curves over six years of 
observations (from 1999 to 2004) for V819~Tau with an age of 3.3 Myr 
(a) and for W62 (RXJ0452.5+1730) with an age of 22 Myr (b).}}
\end{figure}

In our previous papers, we showed that a small
periodicity amplitude suggests a more uniform distribution
of spots over the stellar surface, while a
large amplitude is typical of the case where the spots
are concentrated in one or two high-latitude regions,
i.e., they are distributed highly nonuniformly (see
Grankin 1999; Grankin et al. 2008). These conclusions
are confirmed by the Doppler mapping of the
surfaces of selected PMS stars. In particular, cool
long-lived high-latitude spots were shown to have
dominated on the surface of V410~Tau in the period
1992–1993, when the photometric variability amplitude
was at a maximum and reached $0^m.5-0^m.6$ (Rice
and Strassmeier 1996). In contrast, in the period
2007–2009, when the photometric variability amplitude
decreased to $0^m.06-0^m.10$ (Grankin and Artemenko
2009), Doppler mapping showed that quite
a few low-latitude spots distributed in longitude almost
uniformly were present on the stellar surface
(see Fig. 4 in Skelly et al. 2010). Thus, it can be
assumed that the above differences in photometric
behavior between the youngest PMS stars and older
sample objects are attributable to different patterns
of distribution of the spots over the stellar surface.
Since the positions of cool magnetic spots on the
surfaces of stars with convective envelopes are related
to the locations of local magnetic fields, it is obvious
that the pattern of distribution of the spots over the
surface will depend on the magnetic field structure.
Since the photometric behavior of the most active and
youngest objects suggests that long-lived spots are
concentrated at high latitudes, it can be assumed that
the magnetic field of these stars has a simpler and
fairly symmetric dipole structure. Owing to such a
structure, the spots are concentrated predominantly
near the magnetic poles and retain their positions
during many rotation cycles. It is such a behavior
that we observe in the case of the most active and
youngest objects. In contrast, the relatively old stars
can have a more complex magnetic field structure.
Therefore, the spots are distributed over the surface
more uniformly; the amplitude of the periodicity is
much smaller or it is not observed at all. Such a
photometric behavior is typical of the older objects
from our sample. In other words, the existence of a
correlation between the photometric behavior and age
can be a consequence of the evolution of the magnetic
field structure for PMS stars.


\begin{figure}[h]
\epsfxsize=14cm
\vspace{0.5cm}
\hspace{1cm}
\epsffile{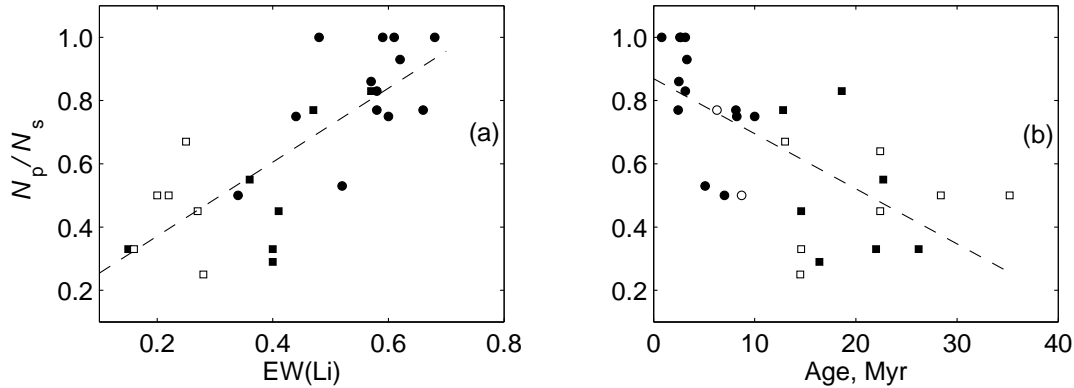}
\caption{\rm \footnotesize {Frequency of occurrence of a periodicity 
$f=N_{\rm p}/N_{\rm s}$ versus lithium line equivalent width (a) and 
age (b).}}
\end{figure}

This assumption is in good agreement with the results of a recent study 
of the magnetic field topology for several PMS stars performed within 
the MaPP (Magnetic Protostars and Planets) Program (see, e.g., Donati 
et al. (2010, 2011) and references therein). In particular, these 
studies showed that the magnetic field structure evolves from predominantly
dipole and axisymmetric (in the case of fully convective stars) to 
octupole and axisymmetric (when the radiative core is less than half 
the stellar radius) and then to multipole and nonaxisymmetric (when
the convective zone occupies less than half the stellar radius). The 
fact that the five most active stars discussed above lie on the 
Hertzsprung–Russell diagram in the region where fully convective PMS stars
with a predominantly dipole and axisymmetric magnetic field structure 
are located can serve as an additional argument for such evolution of 
the magnetic field. Our future cooperative studies of the magnetic 
field topology for 40 PMS stars planned as part of the MaTYSSE 
(Magnetic Topologies of Young Stars \& the Survival of close-in massive 
Exoplanets) Program performed on TBL (NARVAL spectrograph) and CFHT 
(ESPaDOnS spectrograph) will show whether the changes in the amplitude 
of the light curve are accompanied by significant changes in the
distribution of spots and/or magnetic field topology.

\subsection*{CONCLUSIONS}
\indent

We analyzed a sample of 70 magnetically active
stars toward the Taurus–Auriga SFR and investigated
the relationship between magnetic activity and
rotation for these objects. In particular, we obtained
the following results.

We analyzed the relationship between various X-ray
activity parameters and rotation for PMS stars
in the Taurus–Auriga SFR. We showed that there
is no significant correlation between various X-ray
activity parameters ($L_{\rm X}$, $L_{\rm X}/L_{\rm bol}$, and 
$F_{\rm X}$) and rotation
parameters, such as the period and the angular
momentum. We investigated the positions of
PMS stars on the Rossby diagram. All sample stars
exhibit the same $L_{\rm X}/L_{\rm bol}$ and $R_0$ as the stars 
from the Pleiades and IC 2602 clusters with ages within the
range $30-100$ Myr, i.e., they are in the regime of a
saturated dynamo.

We analyzed the photospheric activity of PMS stars
based on original long-term photometric observations.
The maximum photometric variability amplitude
was found, on average, to decrease with
increasing age of the sample objects and to increase
with increasing equivalent width of the $\rm H\alpha$ emission
line and the lithium absorption line.

We found a statistically significant correlation between
the lithium line equivalent width and the age
of solar-mass (in the range from 0.7 to 1.2\,$M_\odot$)
PMS stars. The older the age, the smaller the Li line
equivalent width. This result is in excellent agreement
with the predictions of the classical models explaining
the evolution of the atmospheric Li abundance during
the PMS stage of evolution of solar-mass stars.

We identified a group of five most active PMS stars
that exhibit maximum photometric variability amplitudes
reaching $0.^m4-0.^m8$. All these stars have very
similar physical parameters: spectral types (K6--K7),
rotation periods (2.4--5.7 days), radii (1.3--1.75\,$R_\odot$),
masses (0.74--0.92\,$M_\odot$), and ages (2.5--8.2 Myr).
In addition, they show a prominent emission in $\rm H\alpha$
(from --0.5\,\AA~ to --4.0\,\AA) and the strongest lithium
absorption line (EW(Li) \textgreater 0.57\,\AA).

The most interesting feature of the photometric
behavior of these active stars is related to the stability
of the phase light curve over several observing seasons.
The long-termstability of the phase light curves
manifests itself in the fact that the phase of minimum
light ($\varphi_{\rm min}$) retains its value in the interval 
from 5 to 19 years. Such a feature of the photometric behavior
may be attributable to peculiarities of the magnetic
field configuration for these stars. The most active
and youngest stars from our sample most likely have
mainly large-scale magnetic fields with an axisymmetric
poloidal configuration. In this case, extended
spotted regions are concentrated near the locations of
two magnetic poles. The long existence of extended
spotted regions suggests that the structure of these
dipole fields is fairly stable over several years.

The remaining sample stars exhibit small photometric
variability amplitudes (no more than $0^m.15$),
with a periodicity being observed not in each observing
season. The frequency of occurrence of a
periodicity was shown to be maximal for the youngest
stars and to gradually decrease for older objects. It
may well be that the existence of this relationship
is an indirect confirmation of the evolution of the
magnetic field structure for young stars from predominantly
dipole and axisymmetric (in the case of fully
convective stars) to octupole and axisymmetric (when
the radiative core is less than half the stellar radius)
and then to multipole and nonaxisymmetric (when
the convective zone occupies less than half the stellar
radius). The fact that the five most active stars lie
on the Hertzsprung–Russell diagram in the region
where fully convective PMS stars with a fairly simple
dipolar magnetic field structure are located can serve
as an additional argument for such evolution of the
magnetic field.


\subsection*{REFERENCES}

\begin{enumerate}

\item S.A. Barnes, Astrophys. J. {\bf 561}, 1095 (2001).

\vspace{-1ex}
\item J.-F. Donati, M.B. Skelly, J. Bouvier, et al.,
Mon. Not. R. Astron. Soc. {\bf 409}, 1347 (2010).

\vspace{-1ex}
\item J.-F. Donati, J. Bouvier, F.M. Walter, et al.,
Mon. Not. R. Astron. Soc. {\bf 412}, 2454 (2011).

\vspace{-1ex}
\item K.N. Grankin, Astron. Lett. 
{\bf 25}, 526 (1999). 

\vspace{-1ex}
\item K.N. Grankin and S.A. Artemenko,
Inform. Bull. Var. Stars,  5907 (2009).

\vspace{-1ex}
\item K.N. Grankin, Astron. Lett. {\bf39}, 251 (2013a).

\vspace{-1ex}
\item K.N. Grankin, Astron. Lett. {\bf39}, 336 (2013b).

\vspace{-1ex}
\item K.N. Grankin, J. Bouvier, W. Herbst, et al., Astron.
Astrophys. {\bf 479}, 827 (2008). 

\vspace{-1ex}
\item K. N. Grankin, M. A. Ibragimov, V. B. Kondrat’ev, et
al., Astron. Rep. {\bf 39}, 799 (1995). 

\vspace{-1ex}
\item A. Hempelmann, J.H.M.M. Schmitt, M. Schultz, et al.,
Astron. Astrophys. {\bf 294}, 515 (1995).

\vspace{-1ex}
\item R.D. Jeffries, in {\it Chemical Abundances and Mixing in
Stars in the Milky Way and its Satellites}, Ed. by
S. Randich and L. Pasquini (Springer, Berlin, 2006), p. 163.

\vspace{-1ex}
\item R.W. Noyes, L.W. Hartmann, S.L. Baliunas, et al., 
Astrophys. J. {\bf 279}, 763 (1984).

\vspace{-1ex}
\item B.M. Patten and T. Simon,
Astrophys. J. Suppl. Ser. {\bf 106}, 489 (1996).

\vspace{-1ex}
\item N. Pizzolato, A. Maggio, G. Micela, et al.,
Astron. Astrophys. {\bf 397}, 147 (2003).

\vspace{-1ex}
\item D. Queloz, S. Allain, J.-C. Mermilliod, et al.,
Astron. Astrophys. {\bf 335}, 183 (1998).

\vspace{-1ex}
\item S. Randich, J.H.M.M. Schmitt, C.F. Prosser, et al.,
Astron. Astrophys. {\bf 305}, 785 (1996).

\vspace{-1ex}
\item J.B. Rice and K.G. Strassmeier, 
Astron. Astrophys. {\bf 316}, 164 (1996).

\vspace{-1ex}
\item A. Scholz, J. Coffey, A. Brandeker, et al., 
Astrophys. J. {\bf 662}, 1254 (2007).

\vspace{-1ex}
\item C. J. Schrijver and C. Zwaan, 
{\it Solar and Stellar Magnetic Activity}
(Cambridge Univ. Press, Cambridge,
2000).

\vspace{-1ex}
\item M.B. Skelly, J.-F. Donati, J. Bouvier, K.N. Grankin, et al.,
Mon. Not. R. Astron. Soc. {\bf 403}, 159 (2010).

\vspace{-1ex}
\item A. Skumanich,
Astrophys. J. {\bf 171}, 565 (1972).

\vspace{-1ex}
\item D.M. Terndrup, J.R. Stauffer, M.H. Pinsonneault, et al.,
Astron. J. {\bf 119}, 1303 (2000).

\vspace{-1ex}
\item R. Wichmann, J. Krautter, J.H.M.M. Schmitt, et al., 
Astron. Astrophys. {\bf 312}, 439 (1996).

\end{enumerate}

\end{document}